\documentclass[preprintnumbers,nofootinbib,noshowpacs,eqsecnum,superscriptaddress,prd]{revtex4}
\pdfoutput=1
\usepackage{graphicx}
\usepackage{amsmath,amssymb,mathrsfs}
\usepackage{fancyhdr}
\usepackage{psfrag}
\usepackage{array,bbm} 
\usepackage{url}

\newcommand*{\tvec}[1]{\ensuremath{\boldsymbol{\mathrm{#1}}}}           

\makeatletter       

\renewcommand{\p@subsection}{}

\makeatother

\DeclareMathOperator{\im}{Im}

\begin{document}
\title{
The MCPM, a two-Higgs-doublet model with maximal CP symmetry, and LHC13}

\author{M. Maniatis}
    \email[E-mail: ]{maniatis8@gmail.com}
    
\affiliation{Departamento de Ciencias B\'a{}sicas, 
UBB, Casilla 447, Chill\'a{}n, Chile.}

\author{O. Nachtmann}
    \email[E-mail: ]{O.Nachtmann@thphys.uni-heidelberg.de}

\affiliation{
Institut f\"ur Theoretische Physik, Philosophenweg 16, 69120
Heidelberg, Germany
}

\begin{abstract}
We continue our investigation of the phenomenological consequences of the
MCPM for the LHC experiments. As in any two-Higgs-doublet model we have in 
the MCPM three neutral Higgs bosons and one charged Higgs-boson pair $H^\pm$.
Here we discuss the two-photon production in proton-proton collisions.
We find that in the MCPM a resonance-type structure in the $\gamma\gamma$ 
invariant mass distribution is predicted around twice the $H^\pm$ mass $m_{H^\pm}$
with a width $2\; \Gamma_H$ where $\Gamma_H$ is the $H^\pm$ width.
If we set $m_{H^\pm} = 375$ GeV, the above resonance structure appears at 750 GeV with
a width of about 45 GeV. We point out various predictions of the MCPM which
follow in such a scenario and which can be checked at the LHC.
\end{abstract}

\maketitle

%
\section{Introduction}
\label{sec-intro}

One of the aims of the present LHC experiments is the exploration
of the scalar sector of particle physics. Indeed, one scalar particle
was already found in brilliant experiments \cite{Aad:2012tfa,Chatrchyan:2012xdj}.
But it is not clear if Nature corresponds to the Standard Model (SM) where we 
have only one physical Higgs boson. Many models with extended scalar 
sectors exist in the literature. An attractive alternative to the SM
are two-Higgs-doublet models, THDMs; see \cite{Gunion:1989we,Branco:2011iw}
and references therein. 
In our group we emphasised the usefulness of bilinears for
the study of THDMs 
\cite{Nagel:2004sw,Maniatis:2006fs,Maniatis:2006jd,Maniatis:2007vn,Maniatis:2007de}.
A THDM with maximal CP symmetry has been presented in \cite{Maniatis:2007de}.
This model, the {\em maximal CP symmetric model}, MCPM, gives a certain understanding
of family replication and fermion mass hierarchies.
Phenomenological consequences of the MCPM were worked out
in \cite{Maniatis:2009vp,Maniatis:2009by,Maniatis:2010sb,Maniatis:2011qu,Brehmer:2012hh}.
In the present paper we continue the phenomenological investigations of the MCPM
in view of the possibilities of the experiments at LHC13.
We shall, in particular, be interested in two-photon production in proton-proton
collisions.

Our paper is organised as follows. In section \ref{sec-MCPM} we 
briefly review some features of the MCPM. In section 
\ref{sec-pp2ggX} we  present the details of the calculation of
diphoton production in proton-proton collisions.
Section \ref{sec-discussion} contains our discussion and section \ref{sec-conclusions}
our conclusions.

%
\section{Brief review of the MCPM}
\label{sec-MCPM}

In this section we recall briefly some main features of the MCPM.
Like in any two-Higgs-doublet model there are five physical Higgs bosons,
in our notation, $\rho'$, $H^\pm$, $h''$, $h'$. 
The $\rho'$ behaves on its mass shell very similarly to the SM Higgs boson
which we denote by $\rho'_{\text{SM}}$. Thus, we set for the mass of the
$\rho'$ the measured value from \cite{Aad:2012tfa,Chatrchyan:2012xdj}.
\begin{equation} \label{2.1}
m_{\rho'} = 125 \text{ GeV}.
\end{equation}
The masses of the charged Higgs-boson pair, $H^\pm$, of the
pseudoscalar $h''$ and of the scalar $h'$ are not predicted 
by the model except that the hierarchy
\begin{equation}
m_{h''} < m_{h'}
\end{equation}
is required. The model is built to have the generalised CP symmetry
of type (i), see \cite{Maniatis:2007de}, and this has drastic consequences.
The couplings of the Higgs bosons among themselves are determined in terms of 
the masses. Furthermore, in the strict symmetry limit and concerning the
Yukawa couplings, the third fermion family couples exclusively to the $\rho'$, the
second family only to $H^\pm$, $h''$, $h'$, but with 
coupling constants related to the third family.
The first family of fermions is uncoupled to the Higgs sector in this limit
where the masses of the second and first-family fermions are zero.
Of course, this is not realistic, but it is not so bad as a first
approximation to Nature; see \cite{Maniatis:2007de}.
In the following we will work in this strict symmetry limit. Thus,
all relations are subject to corrections from symmetry breaking.
But in this paper we are interested in the physics of the Higgs bosons of 
the MCPM, that is, physics at the 100 to 800 GeV scale. We expect
that symmetry breaking corrections due to the non-zero masses of the first and second
fermion family will be small in this regime.
The Lagrangian of the MCPM is given explicitly in appendix A of \cite{Maniatis:2009vp}.

\section{The MCPM and the reaction $pp\to \gamma\gamma X$}
\label{sec-pp2ggX}

\begin{figure}[t] 
\centering
\includegraphics[width=0.7\linewidth,clip]{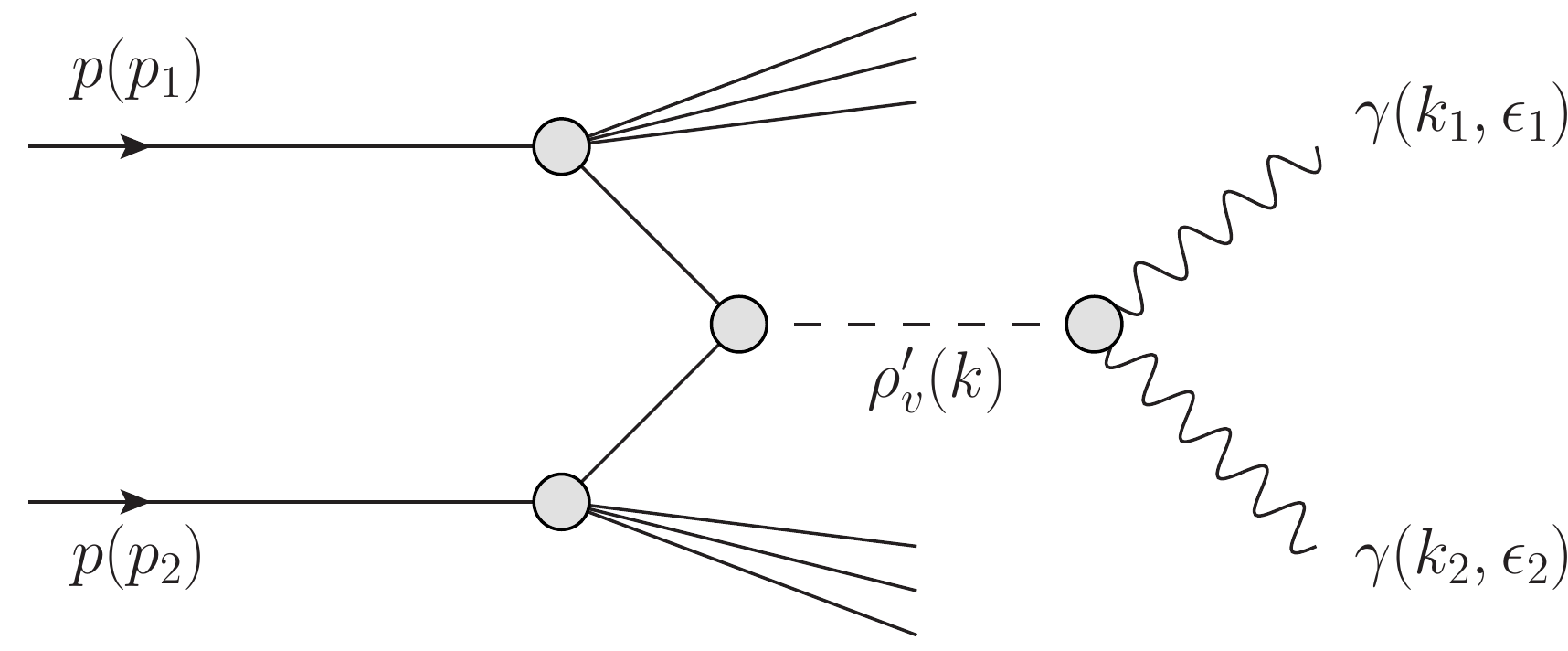}
\caption{\label{pp2gg}
Diphoton production in proton-proton collision via an intermediate virtual
Higgs boson $\rho_v'$.} 
\end{figure}

In the following we shall study the reaction of two protons, $p(p_1)$ and $p(p_2)$, 
giving a pair of photons, $\gamma(k_1, \epsilon_1)$ 
and $\gamma(k_2, \epsilon_2)$,
via an intermediate virtual Higgs boson $\rho_v'(k)$, and a rest $X$, 
see Fig. \ref{pp2gg},
\begin{equation} \label{reaction}
p(p_1) + p(p_2) \to \rho'_v(k) + X \to \gamma(k_1, \epsilon_1)+ \gamma(k_2, \epsilon_2) + X.
\end{equation}
Here the 
momentum and polarization vectors are indicated in brackets.
The ${\bf T}$-matrix element for this process is
\begin{multline} \label{timeordered}
i \langle \gamma(k_1, \epsilon_1), \gamma(k_2, \epsilon_2), X |{\bf T}| p(p_1), p(p_2) \rangle
=\\
i \langle \gamma(k_1, \epsilon_1), \gamma(k_2, \epsilon_2) |{\bf T}| \rho_v'(k) \rangle
\frac{i}{k^2-m_{\rho'}^2 + i m_{\rho'} \Gamma_{\rho'}}
i \langle \rho_v'(k), X |{\bf T}| p(p_1), p(p_2) \rangle,
\end{multline}
with $k=k_1+k_2$, and $m_{\rho'}$, $\Gamma_{\rho'}$ the mass and width of the
Higgs boson $\rho'$, respectively.
The invariant mass squared of
the photon pair and the square of the center-of-mass collision energy are defined as usual,
\begin{equation}
m_{\gamma \gamma}^2 = (k_1 + k_2)^2, \qquad
s = (p_1+p_2)^2.
\end{equation}
For unpolarized protons as well as unobserved polarizations of the photons we 
find the cross section
\begin{multline} \label{2.6}
\frac{d \sigma}{d m_{\gamma \gamma}^2} (p(p_1) + p(p_2) \to \gamma(k_1)+ \gamma(k_2) + X(p_X))
=\\
\frac{1}{2 \sqrt{s (s-4 m_p^2)}}
\frac{1}{2}
\int \frac{ d^4k_1}{(2\pi)^3} \delta_+(k_1^2)
\int \frac{ d^4k_2}{(2\pi)^3} \delta_+(k_2^2)  \\
\times \sum_X
(2 \pi)^4 \delta^{(4)}(k_1+k_2+p_X-p_1-p_2)\;
\delta_+( (k_1+k_2)^2 - m_{\gamma \gamma}^2 ) \\
\times \sum_{\text{spins}} \left| \langle \gamma(k_1,\epsilon_1), \gamma(k_2, \epsilon_2)|{\bf T}|\rho_v'(k)\rangle \right|^2
\left|m_{\gamma \gamma}^2 - m_{\rho'}^2 + i m_{\rho'} \Gamma_{\rho'} \right|^{-2} \\
\times \frac{1}{4}
\sum_{\text{spins}} \left|\langle \rho_v'(k), X(p_X) |{\bf T}| p(p_1), p(p_2) \rangle \right|^2.  
\end{multline}
Now we define the production cross section for the virtual Higgs boson $\rho'_v$
\begin{multline} \label{2.7}
\sigma(p(p_1) + p(p_2) \to \rho_v'(m_{\gamma \gamma}^2) +X) = 
\frac{1}{2 \sqrt{s (s-4 m_p^2)}} 
\sum_X \int \frac{ d^4k}{(2\pi)^3} \delta_+(k^2-m_{\gamma \gamma}^2)\; \\
\times (2 \pi)^4 \delta^{(4)}(k+p_X-p_1-p_2)\;
\frac{1}{4}
\sum_{\text{spins}} \left|\langle \rho_v'(k), X(p_X) |{\bf T}| p(p_1), p(p_2) \rangle \right|^2 .
\end{multline}
We define the decay width of the virtual boson $\rho'_v$ of mass squared $m_{\gamma\gamma}^2$ as
\begin{multline}
\Gamma(\rho_v'(m_{\gamma\gamma}^2) \to \gamma \gamma) =
\frac{1}{2 m_{\gamma \gamma}} \frac{1}{2}
\int \frac{ d^4k_1}{(2\pi)^3} \delta_+(k_1^2)
\int \frac{ d^4k_2}{(2\pi)^3} \delta_+(k_2^2)\\
\times (2 \pi)^4 \delta^{(4)}(k_1+k_2-k)\;
\sum_{\text{spins}} \left| \langle \gamma(k_1,\epsilon_1), \gamma(k_2, \epsilon_2)|{\bf T}|\rho_v'(k)\rangle \right|^2.
\end{multline}
With this we have for the  cross sections
\begin{multline} \label{22.9}
\frac{d \sigma}{d m_{\gamma \gamma}^2} (p(p_1) + p(p_2) \to \gamma(k_1)+ \gamma(k_2) + X)
=\\
\sigma(p(p_1) + p(p_2) \to \rho_v'(m_{\gamma \gamma}^2) +X) 
\frac{m_{\gamma \gamma}}{\pi} 
\left|m_{\gamma \gamma}^2 - m_{\rho'}^2 + i m_{\rho'} \Gamma_{\rho'} \right|^{-2}
\Gamma(\rho_v'(m_{\gamma\gamma}^2) \to \gamma \gamma)
\end{multline}
and
\begin{multline} \label{22.10}
\frac{d \sigma}{d m_{\gamma \gamma}} (p(p_1) + p(p_2) \to \gamma(k_1)+ \gamma(k_2) + X)
=\\
\sigma(p(p_1) + p(p_2) \to \rho_v'(m_{\gamma \gamma}^2) +X)
\frac{2 m_{\gamma \gamma}^2}{\pi} 
\left|m_{\gamma \gamma}^2 - m_{\rho'}^2 + i m_{\rho'} \Gamma_{\rho'} \right|^{-2} 
\Gamma(\rho_v'(m_{\gamma\gamma}^2) \to \gamma \gamma).
\end{multline}

\subsection{Production of $\rho'_v$}
\label{sec-prod}

The couplings of the Higgs boson $\rho'$ to the gauge bosons and fermions
of the third generation are like those for the SM Higgs boson $\rho'_{\text SM}$. 
For the production reaction
\begin{equation} \label{prod}
p(p_1) + p(p_2) \to \rho'_v(k) + X
\end{equation}
we have, therefore, the following main processes:
\begin{itemize}
\item gluon-gluon fusion,
\begin{equation}
G+G \to \rho'_v,
\end{equation}
\item vector-boson fusion,
\begin{equation}
W^+ + W^- \to \rho'_v,
\qquad
Z + Z \to \rho'_v,
\end{equation}
\item fusion of $t\bar{t}$ and $b\bar{b}$ quarks,
\begin{equation} \label{tbprod}
t + \bar{t} \to \rho'_v,
\qquad
b + \bar{b} \to \rho'_v.
\end{equation}
\end{itemize}
Therefore, the production cross section \eqref{2.7} can be calculated
as for a SM Higgs boson $\rho'_{\text SM}$ of mass $m_{\gamma\gamma}$
but we have to leave out
very small contributions to $\rho'_{\text SM}$ production
from the annihilation of first and second generation quark-antiquark
pairs. All this has already been discussed in \cite{Maniatis:2009vp,Maniatis:2009by} and
we can thus rely on the results presented there for the cross section \eqref{2.7}.

\subsection{The decay $\rho'_v \to \gamma \gamma$}
\label{sec-decay}

For the decay of the virtual Higgs boson $\rho'_v$ to two photons we have
in the MCPM contributions from fermion loops, $W^\pm$-boson loops,
and $H^\pm$ loops. The latter contributions will be of particular
interest for us in the following.
From gauge invariance and Bose symmetry we can write the amplitude for
\begin{equation} \label{2.15}
\rho'_v(k) \to \gamma(k_1, \epsilon_1) + \gamma(k_2, \epsilon_2)
\end{equation}
as follows
\begin{equation} \label{2.16}
\begin{split}
&
\langle \gamma(k_1, \epsilon_1), \gamma(k_2, \epsilon_2) |{\bf T}|  \rho_v'(k) \rangle
= e^2 \epsilon_1^{\mu *} \epsilon_2^{\nu *} T_{\mu \nu}(k_1, k_2),\\
&
T_{\mu \nu}(k_1, k_2) = \big[ - (k_1 k_2) g_{\mu \nu}+ k_{2\mu} k_{1\nu}+k_{1\mu} k_{2\nu} \big] T(k^2).
\end{split}
\end{equation}
Here $T(k^2)$ is a scalar function receiving contributions from the above mentioned loops which we discuss
now in turn.
\begin{itemize}
\item Fermion loops

The $\rho'_v$ couples to the third generation fermions $t$, $b$, $\tau$ like the SM Higgs
boson $\rho'_{\text SM}$; see Fig. \ref{fermionloop}.
\begin{figure}[t] 
\centering
\includegraphics[width=0.3\linewidth]{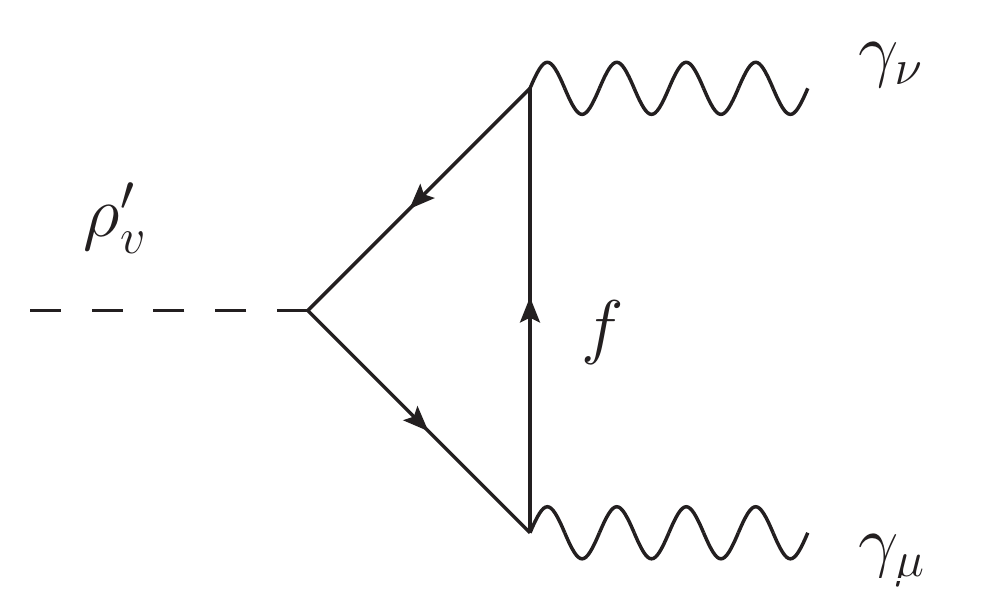}
\quad
\includegraphics[width=0.3\linewidth]{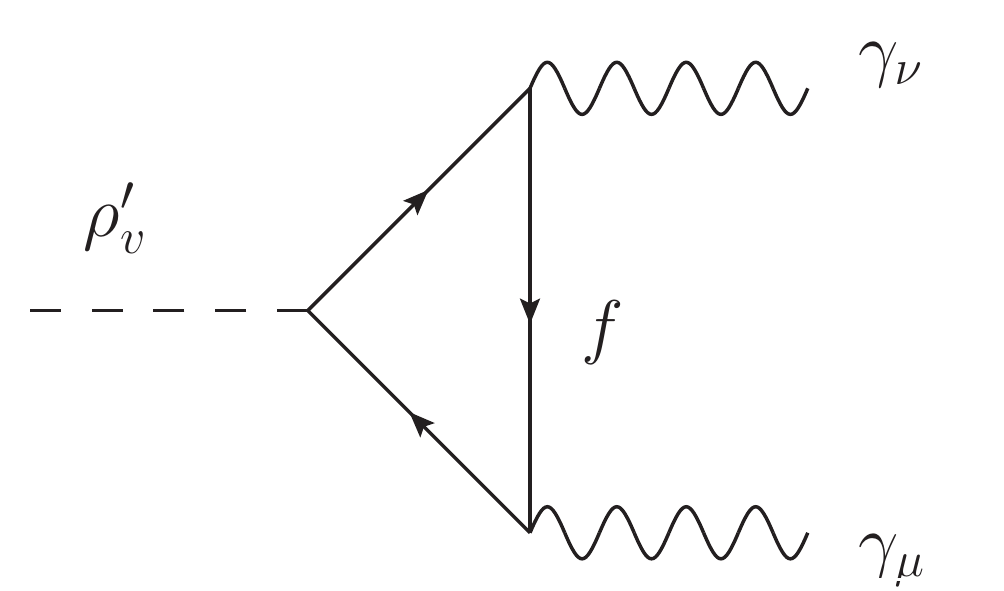}
\caption{\label{fermionloop}
Fermion loop contributions to $\rho'_v \to \gamma \gamma$ for $f=t$, $b$, $\tau$.}
\end{figure}
The calculation of this contribution to $T(k^2)$ in \eqref{2.16}
is standard (see e.g. \cite{Gunion:1989we, Maniatis:2009vp})
and gives, with the couplings as specified in appendix A of 
\cite{Maniatis:2009vp},
\begin{equation} \label{2.17}
T_f(k^2) = - \frac{1}{8\pi^2 v_0}
4 N_c^f e_f^2 \frac{m_f^2}{k^2}
F_{1/2}^{\rho'}\left(\frac{4 m_f^2}{k^2}\right), \qquad f=t,b,\tau .
\end{equation}
Here $v_0=246$ GeV is the standard Higgs vacuum-expectation value, $e_f$ 
is the charge of the fermion $f$ in units of the positron charge, and
$N_c^f$ is the colour factor,
\begin{equation} \label{2.18}
N_c^f = \begin{cases} 3, \text{ for } f=t, b,\\
1, \text{ for } f=\tau. \end{cases}
\end{equation}
The function $F_{1/2}^{\rho'}(z)$ is given by
\begin{equation} \label{2.19}
F_{1/2}^{\rho'}(z) = -2 \big[ 1 + (1-z) f(z)\big], \qquad
f(z) = 
\begin{cases}
-\frac{1}{4} \big[ \ln \left(\frac{1+\sqrt{1-z}}{1-\sqrt{1-z}}\right) - i\pi \big]^2, \text{ for } 0<z<1,\\
\arcsin^2 \left( \frac{1}{\sqrt{z}} \right), \text{ for } z \ge 1.
\end{cases}
\end{equation}
For $\sqrt{z}$ and $\sqrt{1-z}$ in \eqref{2.19} the positive branches of
the square roots have to be taken.

\item $W^\pm$ loops

The diagrams for this contribution are shown in Fig. \ref{Wloops}.
\begin{figure}[t] 
\centering
\includegraphics[width=0.3\linewidth]{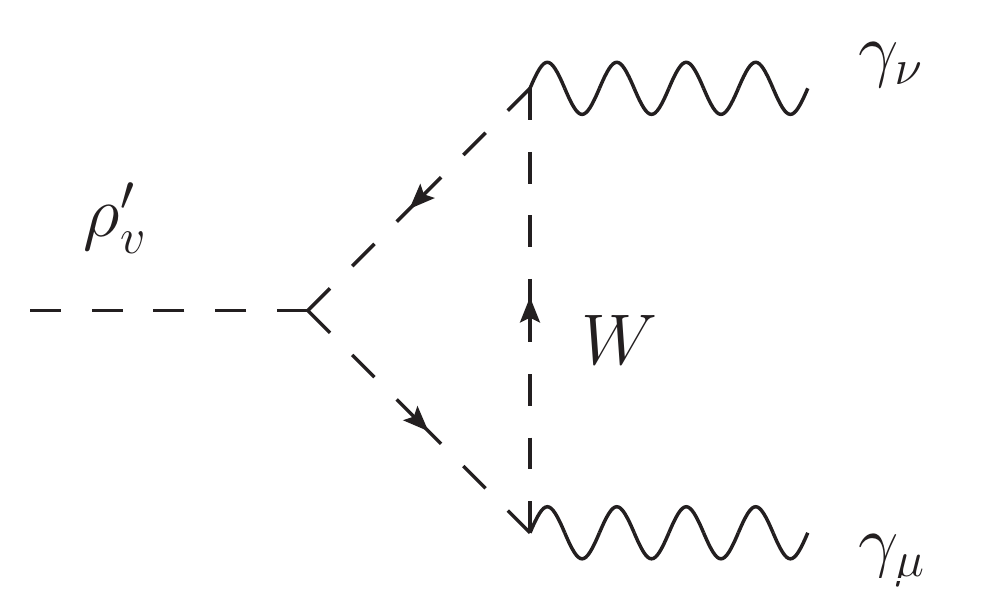}
\quad
\includegraphics[width=0.3\linewidth]{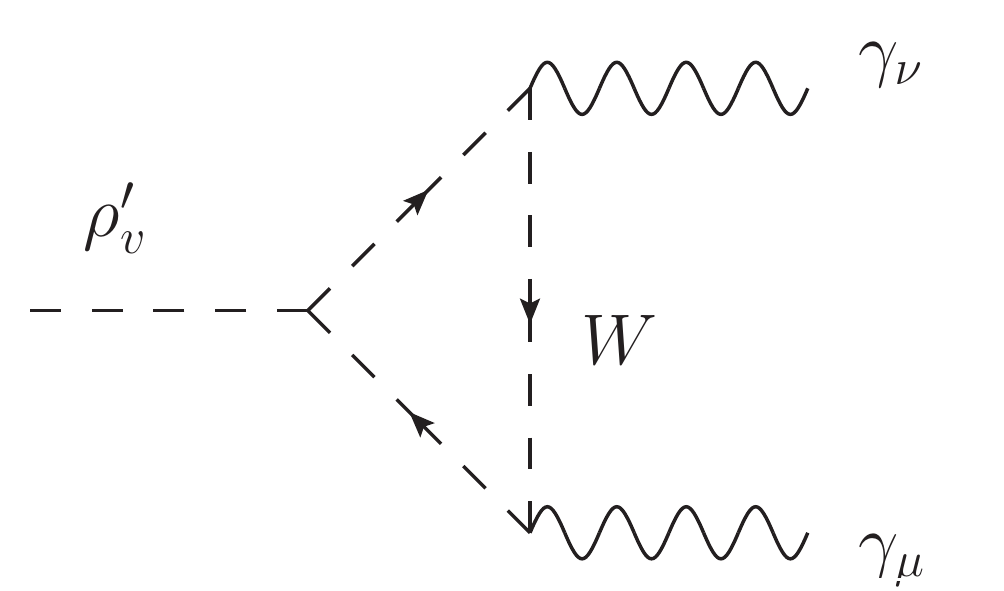}
\quad
\includegraphics[width=0.3\linewidth]{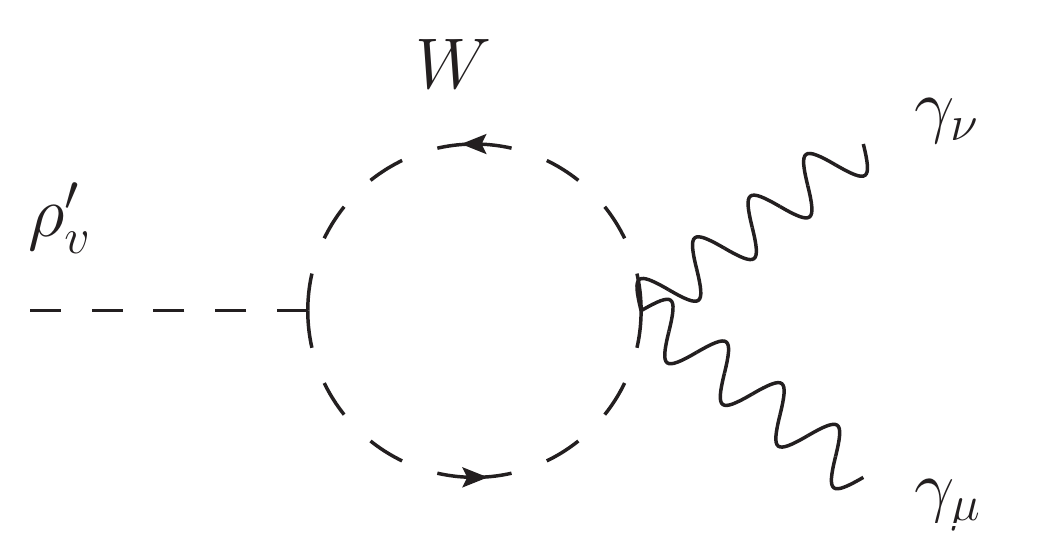}
\caption{\label{Wloops}
$W^\pm$ loop contributions to $\rho'_v \to \gamma \gamma$.}
\end{figure}
Also here the couplings of $\rho'_v$ to $W^\pm$ are as those
for $\rho'_{\text SM}$ to $W^\pm$. Again, we can take over
the standard expressions for this contribution to $T(k^2)$ in \eqref{2.16}
(see \cite{Gunion:1989we, Maniatis:2009vp})
\begin{equation} \label{2.20}
T_W(k^2) = -\frac{1}{8\pi^2 v_0} F_1\left( \frac{4 m_W^2}{k^2} \right),
\quad
F_1(z) = 2 + 3z + 3z(2-z) f(z).
\end{equation}

\item $H^\pm$ loops

The diagrams for this contribution, which has no analogue in the SM, are shown in Fig. 4.
\begin{figure}[t] 
\centering
\includegraphics[width=0.3\linewidth]{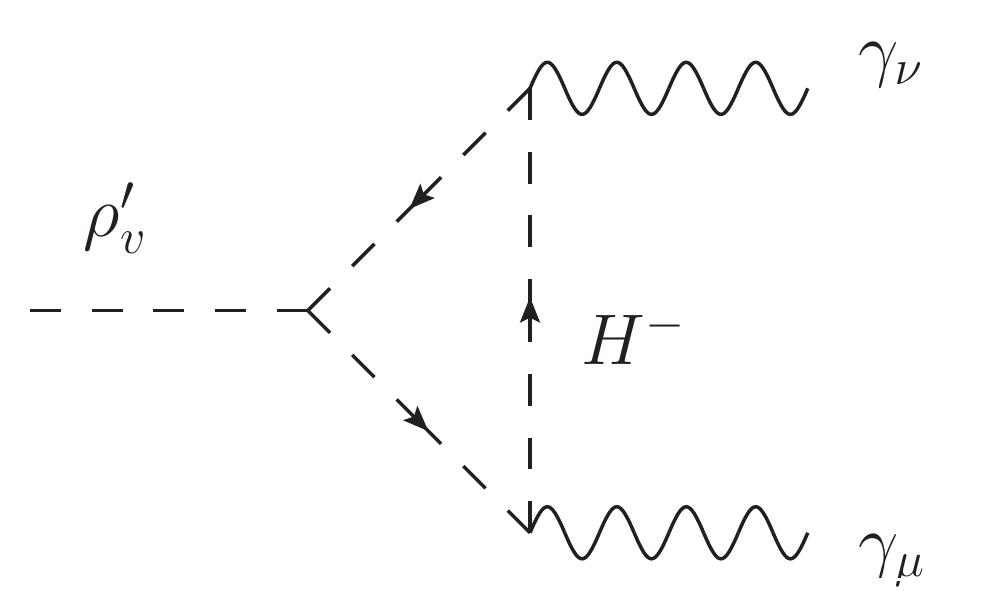}
\quad
\includegraphics[width=0.3\linewidth]{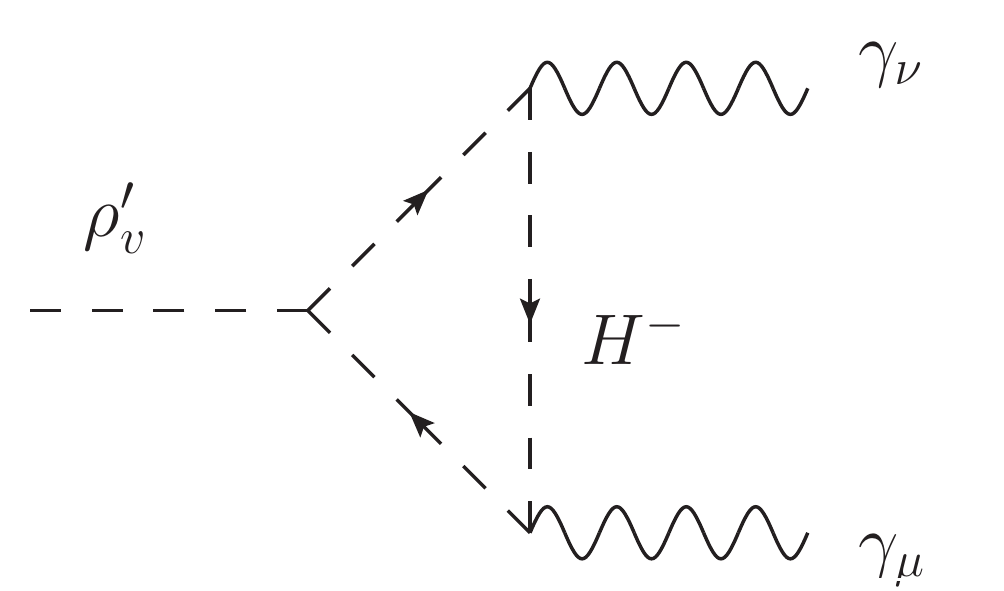}
\quad
\includegraphics[width=0.3\linewidth]{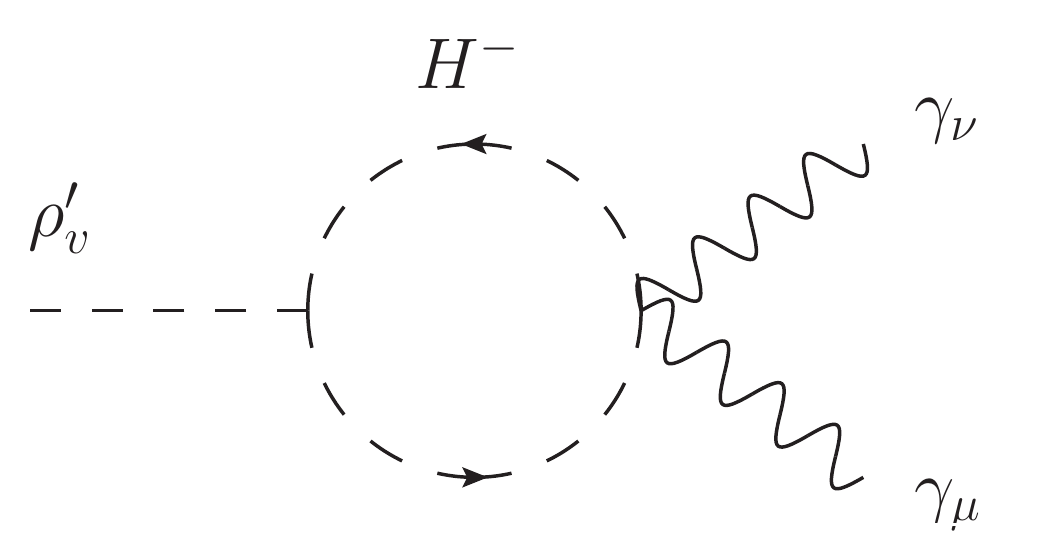}
\caption{\label{Hloops}
$H^\pm$ loop contributions to $\rho'_v \to \gamma \gamma$.}
\end{figure}
Here we encounter the $\rho'H^+H^-$ vertex which is given, in 
the MCPM, as follows (see appendix A of  \cite{Maniatis:2009vp})
\begin{center}
\begin{tabular}[c]{ll} 
\raisebox{-.5\height}{\includegraphics[width=0.3\linewidth]{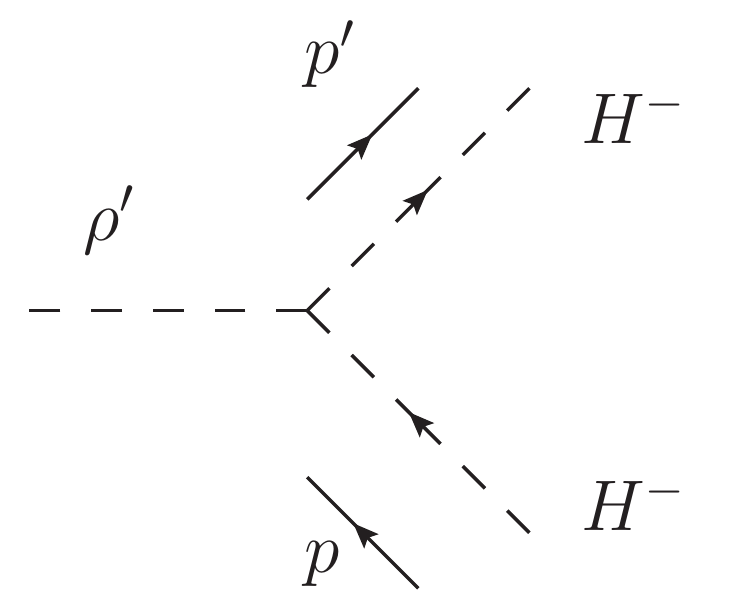}}
&
\qquad
$i \Gamma^{(\rho'HH)}(p',p) = -i \frac{m_{\rho'}^2 + 2 m_{H^\pm}^2}{v_0}$.
\vspace*{\fill}
\end{tabular}
\end{center}

Now, everything is fixed and we get from the diagrams of Fig. \ref{Hloops}
the result
\begin{equation} \label{2.23}
T_{H^\pm}^{(2)} = - \frac{1}{8\pi^2 v_0} \frac{m_{\rho'}^2 + 2 m_{H^\pm}^2}{2 m_{H^\pm}^2}
F_0 \left( \frac{4 m_{H^\pm}^2}{k^2} \right),
\quad \text{where }
F_0(z) = z - z^2 f(z);
\end{equation}
see chapter 3.3 of \cite{Maniatis:2009vp}.
\end{itemize}

\begin{figure}[t] 
\centering
\includegraphics[width=0.25\linewidth]{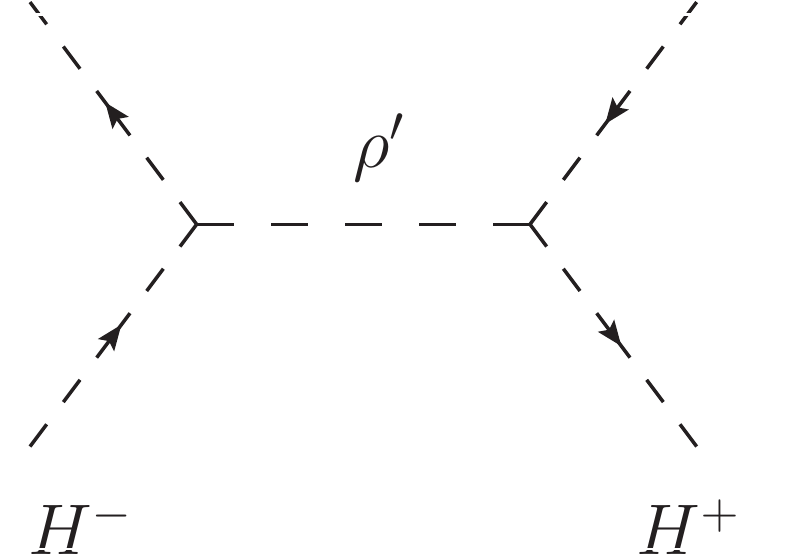}
\quad
\includegraphics[width=0.25\linewidth]{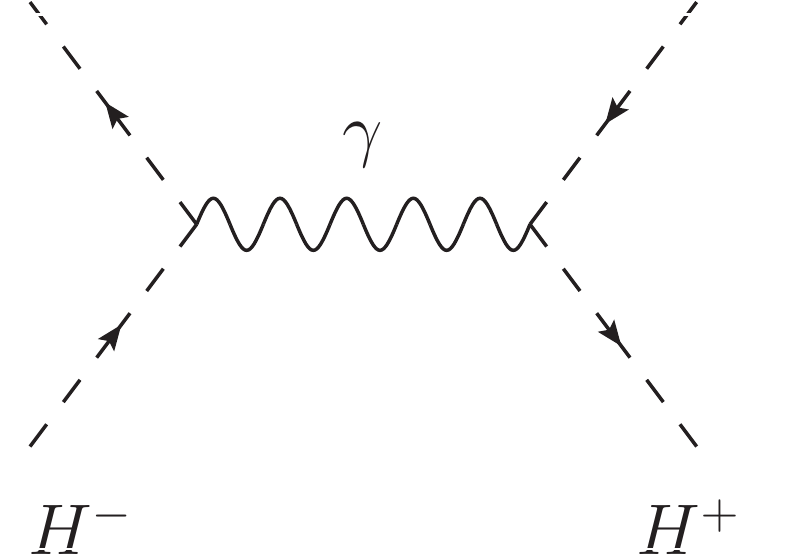}
\caption{\label{pot}
Diagrams of $\rho'$ and $\gamma$ exchange between $H^-$ 
and $H^+$ leading to the potential \eqref{2.26}.}
\end{figure}

\begin{figure}[t] 
\centering
\includegraphics[width=0.4\linewidth]{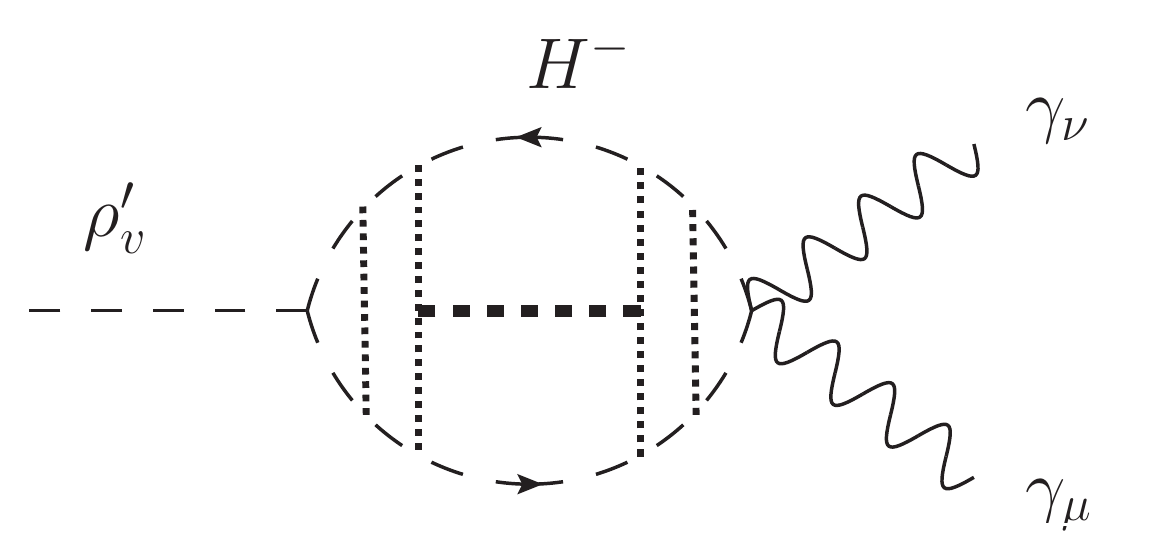}
\caption{\label{Hthreshold}
Diagram of $\rho'_v \to \gamma\gamma$ in the
threshold region \eqref{2.25}. The dotted
lines stand for the exchange due to the potential \eqref{2.26}.}
\end{figure}

But the result \eqref{2.23} is not the whole story.
In the following we shall be mainly interested in the $H^+H^-$ threshold region
\begin{equation} \label{2.25}
k^2 = m_{\gamma\gamma}^2 \approx 4 m_{H^\pm}^2.
\end{equation}
There, as we show now, we have large effects from $H^+ H^-$ interactions.
Indeed, the exchange of $\rho'$ and $\gamma$ between $H^+$ and $H^-$ near threshold
leads to an attractive potential $V(\tvec{x})$ between them, consisting of
a Yukawa and a Coulomb term; see Fig. \ref{pot}.
We get
\begin{equation} \label{2.26}
V(\tvec{x})= - \frac{\kappa}{r} e^{-m_{\rho'} r} - \frac{\alpha}{r},
\end{equation}
where
\begin{equation}
r = |\tvec{x}|, \qquad 
\alpha = \frac{e^2}{4\pi}, \qquad 
\kappa=\frac{1}{4\pi} \frac{1}{4 m_{H^\pm}^2}
\left( \frac{m_{\rho'}^2 + 2 m_{H^\pm}^2}{v_0} \right)^2 .
\end{equation}
To calculate the effects of the potential \eqref{2.26} on 
$\rho'_v \to \gamma\gamma$ in the threshold region \eqref{2.25}
we rely on the methods developed for $t\bar{t}$ production
in its threshold region in \cite{Fadin:1987wz,Fadin:1988fn,Fadin:1990wx,Strassler:1990nw}.
It is easy to see that in the threshold region \eqref{2.25}
the amplitude for the reaction $H^-H^+ \to \gamma\gamma$ is dominated
by the $H^-H^+\gamma\gamma$ contact term. Therefore,
for $\rho'_v \to \gamma\gamma$ in the threshold region
we have to evaluate the diagram shown in Fig. \ref{Hthreshold}.
According to the methods of
\cite{Fadin:1987wz,Fadin:1988fn,Fadin:1990wx,Strassler:1990nw}
we set
\begin{equation} \label{2.28}
\sqrt{k^2} = 2 m_{H^\pm} +E, \qquad
|E| \ll 2 m_{H^\pm}
\end{equation}
and consider the following Green's function $G(\tvec{x},\tvec{y},E)$
defined by the equation
\begin{equation} \label{2.29}
(\hat{H} - i \Gamma_H - E) G(\tvec{x},\tvec{y},E) = \delta^{(3)}(\tvec{x}-\tvec{y}).
\end{equation}
Here $\Gamma_H$ is the width of $H^\pm$ and $\hat{H}$ is the
Hamilton operator
\begin{equation} \label{2.30}
\hat{H} = -\frac{1}{m_{H^\pm}} \Delta_x + V(\tvec{x})
\end{equation}
with $V(\tvec{x})$ from \eqref{2.26} and $\Delta_x$ the Laplace operator.
Suppose then, that the eigenvalue problem for $\hat{H}$ has been solved.
We expect to find discrete eigenvalues
\begin{equation} \label{2.31}
\hat{H} \Psi_{\alpha, \beta}(\tvec{x}) = E_\alpha \Psi_{\alpha, \beta} (\tvec{x})
\end{equation}
for $E_\alpha<0$ labeled by $\alpha=1,2,\ldots$ and possibly a degeneracy
index $\beta$. We normalise the eigenfunctions to 
\begin{equation} \label{2.32}
\int d^3x \Psi^*_{\alpha', \beta'} (\tvec{x})
\Psi_{\alpha, \beta} (\tvec{x}) = \delta_{\alpha',\alpha} \delta_{\beta',\beta}.
\end{equation}
For energies greater or equal to zero we will get a continuous
spectrum
\begin{equation} \label{2.33}
\hat{H} \Psi_{\beta}(\tvec{x},E') = E' \Psi_{\beta}(\tvec{x},E'), 
\qquad E' \ge 0.
\end{equation}
Here, again, $\beta$ is a possible degeneracy label and we normalise
the eigenfunctions to
\begin{equation} \label{2.34}
\int d^3x \Psi^*_{\beta'} (\tvec{x}, E')
\Psi_{\beta''} (\tvec{x},E'') = \delta_{\beta',\beta''} \delta(E'-E'').
\end{equation}
We have then the completeness relation
\begin{equation} \label{2.35}
\sum_{\alpha,\beta}
\Psi_{\alpha, \beta}(\tvec{x})
\Psi^*_{\alpha, \beta} (\tvec{y})
+
\int_0^\infty dE' \sum_\beta
\Psi_{\beta}(\tvec{x},E')
\Psi^*_{\beta}(\tvec{y},E')
= \delta^{(3)}(\tvec{x}-\tvec{y}).
\end{equation}
The Green's function $G(\tvec{x},\tvec{y},E)$ is given by
\begin{equation} \label{2.36}
G(\tvec{x},\tvec{y},E)=
\sum_{\alpha, \beta}
\Psi_{\alpha, \beta}(\tvec{x})
\frac{1}{E_\alpha-E-i \Gamma_H}
\Psi^*_{\alpha, \beta}(\tvec{y})
+
\int_0^\infty dE' \sum_\beta
\Psi_{\beta}(\tvec{x},E')
\frac{1}{E'-E-i \Gamma_H}
\Psi^*_{\beta}(\tvec{y},E').
\end{equation}
Formally the amplitude corresponding to the diagram of Fig. \ref{Hthreshold}
is given by
\begin{equation} \label{2.37}
T_H^{(1)}( 4 m_{H^\pm}^2 + 4 m_{H^\pm} E) =
\frac{1}{4 m_{H^\pm}^4}
\frac{m_{\rho'}^2 + 2 m_{H^\pm}^2}{v_0} G(0,0,E),
\end{equation}
a result which is, however, divergent if we extend
the integration over $E'$ in \eqref{2.36} up to infinity.
But this is not justified since we have to restrict
all energies to be in absolute value much small than $2 m_{H^\pm}$,
the threshold energy. Thus, we introduce a cutoff parameter $E_0$
with
\begin{equation} \label{2.38}
0 < E_0 \ll 2 m_{H^\pm}
\end{equation}
and extend the integral in \eqref{2.36} only up to $E'=E_0$.
We get then from \eqref{2.28} and \eqref{2.36} to \eqref{2.38}
\begin{multline} \label{2.39}
T_H^{(1)}(k^2) =
\frac{1}{m_{H^\pm}^3}
\frac{m_{\rho'}^2 + 2 m_{H^\pm}^2}{v_0} 
\bigg\{
\sum_{\alpha, \text{S waves}}
\frac{|\Psi_\alpha(0)|^2}
{4 m_{H^\pm}^2 + 4 m_{H^\pm} E_\alpha - i 4 m_{H^\pm}\Gamma_H - k^2}
\\
+\int^{E_0}_{0, \text{S waves}} dE' 
\frac{|\Psi(0,E')|^2}
{4 m_{H^\pm}^2 + 4 m_{H^\pm} E' - i 4 m_{H^\pm} \Gamma_{H} - k^2}
\bigg\}.
\end{multline}
Note that only the S waves contribute here 
since we evaluate the Green's function for $\tvec{x}=\tvec{y}=0$.

In order to obtain an (approximate) complete result
for $\rho'_v \to \gamma\gamma$ via the $H^-H^+$ loops we
cannot simply add $T_H^{(1)}(k^2)$, \eqref{2.39}, and
$T_H^{(2)}(k^2)$, \eqref{2.23}.
This would imply a double counting of the threshold region.
We shall thus subtract from $T_H^{(2)}(k^2)$
the contribution from the threshold region.
From the properties of $T_H^{(2)}(k^2)$ we see that
it satisfies an unsubtracted dispersion relation
\begin{equation} \label{2.40}
T_H^{(2)}(k^2) =
\frac{1}{\pi}
\int_{4 m_{H^\pm}^2}^\infty\!\! ds\;
\frac{\im \left(T_H^{(2)}(s) \right)}{s-k^2-i \epsilon}
\end{equation}
with
\begin{equation} \label{2.41}
\im \left(T_H^{(2)}(s)\right)=
\frac{1}{2\pi}
\frac{m_{\rho'}^2 + 2 m_{H^\pm}^2}{v_0 m_{H^\pm}^2} 
\left( \frac{m_{H^\pm}^2}{s} \right)^2
\ln
\left(\frac{1 + \sqrt{1- 4 m_{H^\pm}^2/s}}{1 - \sqrt{1- 4 m_{H^\pm}^2/s}}\right)\;
\theta(s-4 m_{H^\pm}^2).
\end{equation}
We choose now a function $\chi(s,E_0)$ defined for $s \ge 4 m_{H^\pm}^2$
with the properties
\begin{equation} \label{2.42}
\chi(4 m_{H^\pm}^2,E_0) = 1, \quad
\chi(s,E_0) = 0 \text{ for } s \gg 4 m_{H^\pm}^2 + 4 m_{H^\pm} E_0, \quad
\chi(s,E_0) \text{ monotonously decreasing}.
\end{equation}
With this function we set
\begin{equation} \label{2.43}
\im \left(T_H^{(3)}(s) \right)= - \chi(s, E_0) \im \left( T_H^{(2)}(s) \right)
\end{equation}
and
\begin{equation} \label{2.44}
T_H^{(3)}(s)=
\frac{1}{\pi}
\int_{4 m_{H^\pm}^2}^\infty ds
\frac{\im (T_H^{(3)}(s) )}{s-k^2-i \epsilon}.
\end{equation}
We set for the complete contribution of the $H^\pm$ loops
to the function $T(k^2)$ in \eqref{2.16}
\begin{equation} \label{2.45}
T_H(k^2)=T_H^{(1)}(k^2)+T_H^{(2)}(k^2)+ T_H^{(3)}(k^2) .
\end{equation}
By construction $T_H^{(3)}(k^2)$ cancels out the
threshold contribution of $T_H^{(2)}(k^2)$.
In practical calculations we shall choose the function
$\chi(s, E_0)$ \eqref{2.42} such that $\im \left(T_H(k^2)\right)$
has a smooth behaviour in the transition region from threshold
to continuuum.

Putting everything together we have for the function $T(k^2)$ of
\eqref{2.16} from \eqref{2.17}, \eqref{2.20}, and \eqref{2.45}
\begin{equation} \label{2.46}
T(k^2)=
\sum_{f= t, b, \tau}
T_f(k^2)+T_W(k^2)+ T_H(k^2) .
\end{equation}
The decay width of the virtual particle $\rho'_v$ is then
\begin{equation} \label{2.47}
\Gamma(\rho'_v(k^2) \to \gamma\gamma ) =
\frac{\pi}{4} \alpha^2 
\left(k^2\right)^{3/2} 
|T(k^2)|^2 .
\end{equation}.

%
\section{Discussion}
\label{sec-discussion}
 
Looking at the result \eqref{2.39} for $\rho'_v \to \gamma\gamma$ from
the $H^-H^+$ loop in the threshold region we see that it corresponds
to a superposition of resonance contributions. The positions of 
the resonances are approximately at $k^2= 4 m_{H^\pm}^2$, the
widths are $2 \Gamma_H$.
Thus, the MCPM predicts such a type of resonance structure in the $\gamma\gamma$
spectrum. To give a concrete example we shall now choose the mass of
$H^\pm$ to be
\begin{equation} \label{4.1}
m_{H^\pm} = 375 \text{ GeV.}
\end{equation}
Then, we get from section 3.1 of \cite{Maniatis:2009vp} that
the main fermionic decays of $H^\pm$ are
\begin{equation} \label{4.2}
H^- \to s \bar{c}, \qquad H^+ \to c \bar{s}
\end{equation}
giving a decay rate (see table 3 of \cite{Maniatis:2009vp})
\begin{equation} \label{4.3}
\Gamma(H^-\to s\bar{c}) = \Gamma(H^+\to c\bar{s}) = 22.7 \text{ GeV.}
\end{equation}
The decays
\begin{equation} \label{4.3a}
H^-\to h' + W^-, \quad 
H^+\to h' + W^+, \quad 
H^-\to h'' + W^-, \quad 
H^+\to h'' + W^+ 
\end{equation}
can occur in the MCPM if they are energetically possible. But, 
as we will argue below, even then their contribution to the
total width of $H^\pm$ should be small.
Thus, the best estimate for the total width of $H^\pm$ of mass \eqref{4.1}
is
\begin{equation} \label{4.4}
\Gamma_H \equiv \Gamma_{H^-} = \Gamma_{H^+} \approx 22.7 \text{ GeV.}
\end{equation}
The branching fraction of $H^-\to \mu^- \bar{\nu}_\mu$ ($H^+\to \mu^+ \nu_\mu$)
is then estimated to be (see (3.25) of \cite{Maniatis:2009vp})
\begin{equation} \label{4.5}
\frac{\Gamma(H^- \to \mu^- \bar{\nu}_\mu)}{\Gamma_H} =
\frac{\Gamma(H^+ \to \mu^+ \nu_\mu)}{\Gamma_H} 
\approx
3 \times 10^{-5}.
\end{equation}
With such a charged Higgs-boson pair $H^\pm$ and $\rho'$ with masses given
in \eqref{4.1} and \eqref{2.1}, respectively, we get a rather strong attractive potential
\eqref{2.26} with
\begin{equation} \label{4.5a}
\kappa = 0.21 .
\end{equation}
Thus, due to this potential, the MCPM predicts a resonance structure in the $\gamma\gamma$
channel at 
\begin{equation} \label{4.6}
2\; m_{H^\pm} = 750 \text{ GeV}
\end{equation}
with a width 
\begin{equation} \label{4.6a}
2\; \Gamma_H \approx 45.4 \text{ GeV.}
\end{equation}

We note now that the ATLAS and CMS collaborations have indeed reported preliminary
evidence for a sort of resonance structure at
$m_{\gamma\gamma} \approx 750$ GeV with a width of the order of 45 GeV; see 
\cite{atlas,CMS:2016owr}.
Of course, we must be very careful and cannot yet identify the resonance 
structures discussed in section \ref{sec-pp2ggX} of this paper with
this possible experimental finding. 
In any case, we shall have to make a numerical study of the predicted effect.
We shall do this in a separate paper. Here we shall only draw some conclusions based
on the {\em hypothesis} that indeed the structure in the $\gamma\gamma$ spectrum 
at $m_{\gamma\gamma} =$ 750 GeV is real and that it has something to do with
the here discussed threshold effects.

\begin{figure}[t] 
\includegraphics[width=0.4\linewidth]{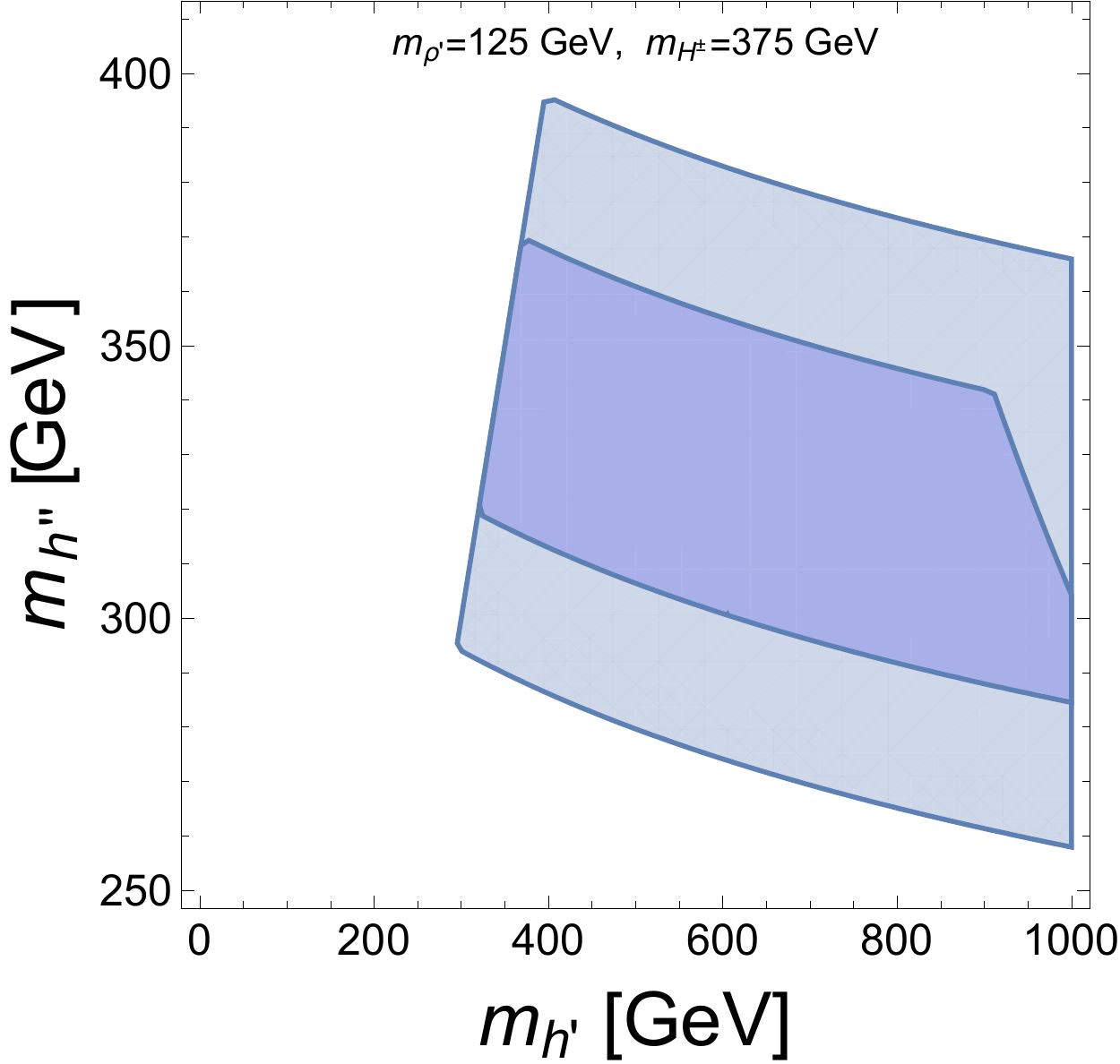}
\caption{\label{oblique}
The allowed regions for the masses of Higgs bosons $h'$ and $h''$
in the MCPM given 
$m_{\rho'} = 125\text{ GeV}$, 
$m_{H^\pm} = 375\text{ GeV}$ and the measured values
of the oblique parameters $S$, $T$, $U$ \cite{Agashe:2014kda}. 
The dark and bright regions correspond to the $1 \sigma$ and 
$2 \sigma$ uncertainties, respectively.}
\end{figure}

The immediate consequence of the above hypothesis is that there should be a charged
Higgs-boson pair $H^\pm$ at roughly $1/2 \times 750 \text{ GeV} = 375 \text{ GeV}$ with
the decay properties discussed in \cite{Maniatis:2009vp} and 
summarised in \eqref{4.2}-\eqref{4.5} here.
We can also say something on the masses of the pseudoscalar ($h''$) and
the scalar ($h'$) Higgs bosons of the MCPM. For this we adapt the analysis of the oblique 
parameters $S$, $T$, $U$ done in \cite{Maniatis:2011qu} for the case
$m_{\rho'} = 125\text{ GeV}$, 
$m_{H^\pm} = 375\text{ GeV}$ (see Fig. 2 of \cite{Maniatis:2011qu}).
The resulting range for $m_{h''}$ versus $m_{h'}$ is shown in Fig. \ref{oblique}.
We see that in the MCPM the masses $m_{h''}$ and $m_{h'}$ 
are predicted to be of the same order, if not higher, than $m_{H^\pm}$.
Thus, the decays \eqref{4.3} should play no role for this mass constellation.
The phenomenology of $h''$ and $h'$ has been discussed extensively in 
\cite{Maniatis:2007de,Maniatis:2009vp,Maniatis:2009by,Maniatis:2010sb,Maniatis:2011qu,Brehmer:2012hh}.
We expect the main production modes to be of the Drell-Yan type
\begin{equation} \label{4.7}
c \bar{c} \to h'', h'
\quad \text{ and } \quad
s \bar{s} \to h'', h' .
\end{equation}
The main decays are predicted to be (see Fig. 8 of \cite{Maniatis:2009vp})
\begin{equation} \label{4.8}
h'' \to c + \bar{c}, \qquad
h'' \to H^\pm + W^\mp \text{ if energetically possible,}
\end{equation}
and
\begin{equation} \label{4.9}
h' \to c + \bar{c}, \qquad
h' \to H^\pm + W^\mp, h'' + Z \text{ if energetically possible.}
\end{equation}
Here we have (see table 3 of \cite{Maniatis:2009vp})
\begin{equation} \label{4.10}
\Gamma(h'' \to c \bar{c}) = 12.08 \text{ GeV} \left( \frac{m_{h'}}{200 \text { GeV}} \right),
\quad
\Gamma(h' \to c \bar{c}) = 12.08 \text{ GeV}  \left( \frac{m_{h''}}{200 \text { GeV}} \right).
\end{equation}
For the decay rates $h'' \to \mu^+\mu^-$ and 
$h' \to \mu^+\mu^-$ the prediction is (see (3.25) of \cite{Maniatis:2009vp})
\begin{equation} \label{4.11}
\frac{\Gamma(h'' \to \mu^-\mu^+)}{\Gamma(h'' \to c\bar{c})} 
\approx
3 \times 10^{-5},
\quad
\frac{\Gamma(h' \to \mu^-\mu^+)}{\Gamma(h' \to c\bar{c})} 
\approx
3 \times 10^{-5}.
\end{equation}

%
\section{Conclusions}
\label{sec-conclusions}

In this article we have discussed the reaction $pp \to \gamma\gamma X$ in the MCPM.
We found that this special two-Higgs-doublet model predicts a 
resonance type structure at $m_{\gamma\gamma} \approx 2 m_{H^\pm}$
with a typical width $2\; \Gamma_H$. 
If this resonance structure in the $\gamma\gamma$ channel is tentatively
put at 750 GeV we predict its width to be around $45.4$ GeV.
Furthermore, we predict $m_{H^\pm} \approx 375$ GeV and
for the pseudoscalar ($h''$) a mass in the range 260 to 400 GeV and 
$m_{h'} > m_{h''}$.
In addition, the MCPM makes definite predictions for the production and decay
of the bosons $H^\pm$, $h''$, $h'$ as discussed in 
\cite{Maniatis:2009vp,Maniatis:2009by,Maniatis:2010sb,Maniatis:2011qu,Brehmer:2012hh}.
A detailed numerical study of the above $\gamma\gamma$ resonance-like structure
will be presented elsewhere.

\section*{Acknowledgments}
The authors are grateful to W. Bernreuther for providing very useful
information concerning the calculation of threshold effects.
This work is supported partly by the 
Chilean research project FONDECYT with project number
1140568 as well as by the group of {\em F\'{i}sica
de Altas Energias} of the UBB, Chile.

\numberwithin{equation}{section}

%
%
%


\begin{thebibliography}{99}

\bibitem{Aad:2012tfa} 
  G.~Aad {\it et al.} [ATLAS Collaboration],
  ``Observation of a new particle in the search for the Standard Model Higgs boson with the ATLAS detector at the LHC,''
  Phys.\ Lett.\ B {\bf 716}, 1 (2012)
  [arXiv:1207.7214 [hep-ex]].

\bibitem{Chatrchyan:2012xdj} 
  S.~Chatrchyan {\it et al.} [CMS Collaboration],
  ``Observation of a new boson at a mass of 125 GeV with the CMS experiment at the LHC,''
  Phys.\ Lett.\ B {\bf 716}, 30 (2012)
  [arXiv:1207.7235 [hep-ex]].

\bibitem{Gunion:1989we} 
  J.~F.~Gunion, H.~E.~Haber, G.~L.~Kane and S.~Dawson,
  ``The Higgs Hunter's Guide,''
  Front.\ Phys.\  {\bf 80}, 1 (2000).

\bibitem{Branco:2011iw} 
  G.~C.~Branco, P.~M.~Ferreira, L.~Lavoura, M.~N.~Rebelo, M.~Sher and J.~P.~Silva,
  ``Theory and phenomenology of two-Higgs-doublet models,''
  Phys.\ Rept.\  {\bf 516}, 1 (2012)
  [arXiv:1106.0034 [hep-ph]].


\bibitem{Nagel:2004sw}
  F.~Nagel,
  ``New aspects of gauge-boson couplings and the Higgs sector,''
PhD-thesis, Heidelberg University (2004),
\mbox{\url{http://www.ub.uni-heidelberg.de/archiv/4803}}


\bibitem{Maniatis:2006fs} 
  M.~Maniatis, A.~von Manteuffel, O.~Nachtmann and F.~Nagel,
  ``Stability and symmetry breaking in the general two-Higgs-doublet model,''
  Eur.\ Phys.\ J.\ C {\bf 48}, 805 (2006)
  [hep-ph/0605184].

\bibitem{Maniatis:2006jd} 
  M.~Maniatis, A.~von Manteuffel and O.~Nachtmann,
  ``Determining the global minimum of Higgs potentials via Groebner bases: Applied to the NMSSM,''
  Eur.\ Phys.\ J.\ C {\bf 49}, 1067 (2007)
  [hep-ph/0608314].
  
\bibitem{Maniatis:2007vn} 
  M.~Maniatis, A.~von Manteuffel and O.~Nachtmann,
  ``CP violation in the general two-Higgs-doublet model: A Geometric view,''
  Eur.\ Phys.\ J.\ C {\bf 57}, 719 (2008)
  [arXiv:0707.3344 [hep-ph]].  

\bibitem{Maniatis:2007de} 
  M.~Maniatis, A.~von Manteuffel and O.~Nachtmann,
  ``A New type of CP symmetry, family replication and fermion mass hierarchies,''
  Eur.\ Phys.\ J.\ C {\bf 57}, 739 (2008)
  [arXiv:0711.3760 [hep-ph]].

\bibitem{Maniatis:2009vp} 
  M.~Maniatis and O.~Nachtmann,
  ``On the phenomenology of a two-Higgs-doublet model with maximal CP symmetry at the LHC,''
  JHEP {\bf 0905}, 028 (2009)
  [arXiv:0901.4341 [hep-ph]].
  
\bibitem{Maniatis:2009by} 
  M.~Maniatis and O.~Nachtmann,
  ``On the phenomenology of a two-Higgs-doublet model with maximal CP symmetry at the LHC. II. Radiative effects,''
  JHEP {\bf 1004}, 027 (2010)
  [arXiv:0912.2727 [hep-ph]].
  
\bibitem{Maniatis:2010sb} 
  M.~Maniatis, O.~Nachtmann and A.~von Manteuffel,
  ``On the phenomenology of a two-Higgs-doublet model with maximal CP symmetry at the LHC: Synopsis and addendum,''
  arXiv:1009.1869 [hep-ph].
  
\bibitem{Maniatis:2011qu} 
  M.~Maniatis and O.~Nachtmann,
  ``Symmetries and renormalisation in two-Higgs-doublet models,''
  JHEP {\bf 1111}, 151 (2011)
  [arXiv:1106.1436 [hep-ph]].

\bibitem{Brehmer:2012hh} 
  J.~Brehmer, V.~Lendermann, M.~Maniatis, O.~Nachtmann, H.-C.~Schultz-Coulon and R.~Stamen,
  ``Towards testing a two-Higgs-doublet model with maximal CP symmetry at the LHC: construction of a Monte Carlo event generator,''
  Eur.\ Phys.\ J.\ C {\bf 73}, no. 4, 2380 (2013)
  [arXiv:1209.2537 [hep-ph]].

\bibitem{Fadin:1987wz} 
  V.~S.~Fadin and V.~A.~Khoze,
  ``Threshold Behavior of Heavy Top Production in e+ e- Collisions,''
  JETP Lett.\  {\bf 46}, 525 (1987)
  [Pisma Zh.\ Eksp.\ Teor.\ Fiz.\  {\bf 46}, 417 (1987)].

\bibitem{Fadin:1988fn} 
  V.~S.~Fadin and V.~A.~Khoze,
  ``Production of a pair of heavy quarks in e+ e- annihilation in the threshold region,''
  Sov.\ J.\ Nucl.\ Phys.\  {\bf 48}, 309 (1988)
  [Yad.\ Fiz.\  {\bf 48}, 487 (1988)].

\bibitem{Fadin:1990wx} 
  V.~S.~Fadin, V.~A.~Khoze and T.~Sj\"ostrand,
  ``On the threshold behaviour of heavy top production,''
  Z.\ Phys.\ C {\bf 48}, 613 (1990).
  
\bibitem{Strassler:1990nw} 
  M.~J.~Strassler and M.~E.~Peskin,
  ``Threshold production of heavy top quarks: QCD and the Higgs boson,''
  Phys.\ Rev.\ D {\bf 43}, 1500 (1991).
    
  
\bibitem{atlas} 
  The ATLAS collaboration,
  ``Search for resonances in diphoton events with the ATLAS detector at $\sqrt{s}$ = 13 TeV,''
  ATLAS-CONF-2016-018, url: \mbox{\url{https://cds.cern.ch/record/2141568}.}
  
\bibitem{CMS:2016owr} 
  CMS Collaboration,
  ``Search for new physics in high mass diphoton events in $3.3~\mathrm{fb}^{-1}$ of proton-proton collisions at $\sqrt{s}=13~\mathrm{TeV}$ and combined interpretation of searches at $8~\mathrm{TeV}$ and $13~\mathrm{TeV}$,''
  CMS-PAS-EXO-16-018, url: \mbox{\url{https://cds.cern.ch/record/2139899}.}

\bibitem{Agashe:2014kda} 
  K.~A.~Olive {\it et al.} [Particle Data Group Collaboration],
  ``Review of Particle Physics,''
  Chin.\ Phys.\ C {\bf 38}, 090001 (2014).
  
  
  
\end{thebibliography}
\end{document}